\documentclass[12pt]{revtex4} 
\usepackage{graphicx} \newcommand{\figspace}{\vspace*{2ex}}
\usepackage{amsmath}
\begin{document}
\titlepage
\title{Wave propagation in a quasi-periodic waveguide network}
\author{Sheelan Sengupta and Arunava Chakrabarti}
\affiliation{Department of Physics, University of Kalyani, Kalyani, West Bengal 741 235, India}
\begin{abstract}
We investigate the transport properties of a classical wave propagating through 
a quasi-periodic Fibonacci array of waveguide segments in the form of loops.
The formulation is general, and applicable for electromagnetic or acoustic waves
through such structures. We examine the conditions for resonant transmission in
a Fibonacci waveguide structure. The local positional correlation between the loops are found to be responsible for the resonance. We also show that, depending on the number of segments attached to a particular loop, the intensity at the nodes displays a perfectly periodic or a self-similar pattern. The former pattern corresponds to a perfectly {\it extended} mode of propagation, which is to be contrasted to the electron or phonon characteristics of a pure one dimensional Fibonacci quasi-crystal.
\end{abstract}
\pacs{42.25.Bs, 42.82.Et, 61.44.Br, 72.15.Rn}
\maketitle
\section{Introduction}
Photonic band gap (PBG) materials have remained an active area of research in condensed matter physics and materials science over the past several years \cite{cms95,sj87,ey87,msk96,dzz94,slm91,dsw97}. With these systems one comes across the interesting possibility of influencing the propagation of electromagnetic wave by creating gaps in the band structure of synthetic periodic dielectric structures \cite{msk96}. Such studies constitute an important part of mesoscopic physics, which focus on the possibility of localization of light \cite{dzz94,slm91}, a direct evidence of which has already been reported for a strongly scattering media of semiconductor powders \cite{dsw97}.
\vskip .2in
Networks formed by segments of one dimensional waveguides provide examples of an alternative class of PBG systems which do not require a material with large dielectric constant. Zhang et al \cite{zqz98} have experimentally observed Anderson localization of light in a three dimensional network, composed of coaxial cables which played the role of a practically one dimensional waveguide. Vasseur et al \cite{jov99} have investigated the photonic band structure of a comb-like waveguide geometry composed of dangling side branches grafted periodically along a mono-mode waveguide. Their work represents a one dimensional photonic crystal enabling one to investigate the occurrence of localized states in such systems. The network system is capable of producing large gaps even in one dimension, though it may be sensitive to the structure of the unit cell.
\vskip .2in
Recently, network models based on serial loop structures (SLS) are being studied in the context of propagation of electromagnetic and acoustic waves \cite{aa03,aa04}. Experiment has been done \cite{aa03} on systems consisting of loop structure made by slender tubes which are pasted on similar tubes of finite length. These circuits allow the propagation of electromagnetic or acoustic wave through practically a single channel. A theory for one dimensional network thus becomes relevant. However, the results presented so far mainly deal with periodic systems, and the influence of a variation from periodicity in the geometry of the network on the transport of classical waves practically remains unaddressed.
\vskip .2in
Motivated by this observation, we undertake a systematic investigation of the transmission of classical waves through a quasi-periodically ordered Fibonacci SLS. We use a model proposed initially by Zhang and Sheng \cite{zqz94} to examine wave propagation in a random array of waveguide segments in the form of loops. Two consecutive loops join each other at a node. Scattering occurs only at the nodes, each pair of which holds an arbitrary number of segments between them. A one dimensional Fibonacci lattice with the golden mean irrationality is already known to exhibit a purely singular continuous spectrum, free from the existence of any extended wave function \cite{mk83}. The interplay of quasi-periodic order and the number of waveguide segments between the consecutive nodes (each segment providing a propagating channel) in a closed loop geometry is likely to produce interesting transmission characteristics, which is the prime objective of the present study. Secondly, whether extended modes of classical wave propagation in a quasi-periodic geometry exists, is yet to be known, to the best of our knowledge. Such extended states do exist \cite{ss93} for an electron traveling through a certain group of quasi-periodic lattices as well as in random lattices with correlated disorder \cite{dhd90,kk93}, where one identifies local clusters of atomic sites, causing resonance and extended wave-function. However, delocalized eigenmodes in the context of quantum transport does not occur in a golden mean Fibonacci chain. It is therefore, interesting to see whether, multiple loops have any role to play in this regard, in the case of classical wave propagation.
\vskip .2in
We find that the band structure of an array of Fibonacci SLS has a strong dependence on the number of waveguide segments between consecutive nodes, which is reflected in the transmission coefficient across a finite Fibonacci SLS. We have also been able to identify local clusters in a Fibonacci SLS, which are responsible for producing resonance leading to finite transmission. The transmission coefficient, at special values of the wave-vector exhibits cyclic variation as a function of the system size. The intensity distribution at the nodes corresponding to such cases may exhibit a completely periodic pattern, or a self-similar (fractal) pattern depending on the mutual relationship between the lengths and the numbers of the waveguide segments in the two basic loop structures constituting the Fibonacci geometry.
\vskip .2in
In what follows, we describe our model, method and the results. A comparison with the corresponding periodic system is made wherever necessary. The formulation remains valid whether we have an electromagnetic wave or sound wave propagating through the network.
\section{The model and the method}
Following Zhang and Sheng \cite{zqz94} we consider a network formed by waveguide segments of flexible length joining together to form a Fibonacci sequence of loops $A$ and $B$ [Fig.1(a)]. A binary Fibonacci sequence is generated from a seed $A$, by using the inflation rule, $A\rightarrow AB$ and $B\rightarrow A$ \cite{mk83}. The `wave-function' $\psi_{i,i+1}$ within any segment of length $l_{i,i+1}$ between the nodes $i$ and $i+1$ is given by \cite{zqz94},
\begin{eqnarray}
\psi_{i,i+1}(x) & = & \psi_i\frac{\sin [k(l_{i,i+1}-x)]}{\sin kl_{i,i+1}}
+\psi_{i+1}\frac{\sin kx}{\sin kl_{i,i+1}}
\end{eqnarray}
\noindent
where, $\psi_i$ is the amplitude at the $i$th mode, and $k$ is the absolute value of the wave vector in the loop. For example, for an electromagnetic wave, $k=\frac{i\omega\sqrt{\epsilon}}{c_0}$ where $\omega$ and $c_0$ are the frequency and the speed (in vacuum) of the electromagnetic wave respectively. $\epsilon$ is the relative permittivity of the dielectric medium, which we may assume to be real. A complex dielectric constant can easily be dealt with. The continuity of the wave function at the nodes and the flux conservation criterion are used to map the problem of wave propagation in such loops into  an equivalent tight binding problem of electron propagation on a one dimension lattice \cite{zqz94}. The resulting difference equation is,
\begin{equation}
(E-\epsilon_i)\psi_i=t_{i,i+1}\psi_{i+1}+t_{i-1,i}\psi_{i-1}
\end{equation} 
\noindent
where, we parametrize $E=2 \cos kl_{i,i+1}$ and, define the effective on-site potential 
\begin{eqnarray}
\epsilon_i & = & 2\cos(kl_{i,i+1})+\sum_{m=1}^{N_{i,i-1}} \cot\theta_{i-1,i}^{(m)}+\sum_{m=1}^{N_{i,i+1}} \cot\theta_{i,i+1}^{(m)}
\end{eqnarray}
The effective nearest neighbour `hopping integral' is given by,
\begin{equation}
t_{i,i+1}=\sum_{m=1}^{N_{i,i+1}}\frac{1}{\sin\theta_{i,i+1}^{(m)}}
\end{equation}
\noindent
Here, $N_{i,i\pm1}$ is the number of segments between the nodes $i$ and $i\pm1$. $\theta_{i,i+1}^{(m)}=kl_{i,i+1}^{(m)}$ is the corresponding phase acquired in the $m^{th}$ segment. The problem of wave propagation now becomes mathematically equivalent to the problem of transmission of an electron with energy $E$ through the mapped lattice, where, $-2 \le E \le 2$. The set of equations (2) is conveniently cast into a matrix form,
\begin{eqnarray}
\left( \begin{array}{c}\psi_{i+1} \\ \psi_{i}
\end{array}\right)
& = & 
\mbox{\boldmath $M$}_i 
\left( \begin{array}{c}\psi_{i} \\ \psi_{i-1}
\end{array}\right)
\end{eqnarray}
\noindent
where,
\begin{displaymath}
\mbox{\boldmath $M$}_i=
 \left( \begin{array}{cc}
\frac{E-\epsilon_i}{t_{i,i+1}} & -\frac{t_{i,i-1}}{t_{i,i+1}}\\ 1 & 0
\end{array}\right)
\end{displaymath}
\noindent
is called the `transfer matrix'.
\vskip .2in
In a Fibonacci SLS of two loops $A$ and $B$, the loops may differ from each other either in terms of the lengths of the segments, or, in the number of segments, or in both. Using the language of electron propagation, on the mapped lattice, which now becomes a one dimensional Fibonacci chain, the `on-site potentials' \cite{mk83} assume three distinct values (in respect of the nearest neighbour loops), viz,
\begin{eqnarray}
\epsilon_\alpha & = & 2 \cos\theta_1 + m[\cot\theta_1+\cot\theta_2]\nonumber\\
\epsilon_\beta & = & 2 \cos\theta_1 + m[\cot\theta_1+\cot\theta_2]+n[\cot\theta_3+\cot\theta_4]\nonumber\\
\epsilon_\gamma & = & \epsilon_\beta
\end{eqnarray}
There are two kinds of `hopping integrals' $t_L$ and $t_S$ for the equivalent electron problem which are arranged accordingly to the Fibonacci sequence. These are given by,
\begin{eqnarray}
t_L & = & \frac{m}{\sin\theta_1}+\frac{m}{\sin\theta_2}\nonumber\\
t_S & = & \frac{n}{\sin\theta_3}+\frac{n}{\sin\theta_4}
\end{eqnarray}
Here, we have assumed that the $A$-loop has $m$ segments of length $a_1$ in the `upper' branch, and $m$ segments of length $a_2$ in the `lower' branch. The $B$-loop has $n$ segments in both branches, the lengths being $a_3$ and $a_4$ for the upper and the lower branches respectively, and $\theta_j=ka_j$ with $a_j=a_1,a_2,a_3,a_4$. The asymmetrical Fibonacci SLS can thus be easily studied. On the equivalent one-dimensional lattice [Fig.1(b)], we define three transfer matrices, viz 
\begin{displaymath}
\mbox{\boldmath $M$}_\alpha
 \left(\begin{array}{cc}
\frac{E-\epsilon_\alpha}{t_L} & -1\\ 1 & 0
\end{array}\right);
\mbox{\boldmath $M$}_\beta 
 \left(\begin{array}{cc}
\frac{E-\epsilon_\beta}{t_S} & -\frac{t_L}{t_S}\\ 1 & 0
\end{array}\right);
\mbox{\boldmath $M$}_\gamma 
 \left(\begin{array}{cc}
\frac{E-\epsilon_\gamma}{t_L} & -\frac{t_S}{t_L}\\ 1 & 0
\end{array}\right)
\end{displaymath}
corresponding to the sites $\alpha$, $\beta$ and $\gamma$ respectively \cite{mk83}. We have examined the spectral properties of both the infinite and the finite Fibonacci SLS for various combinations of $m$, $n$ and $a_j$'s. The fundamental features that we propose to investigate are revealed even if we keep all the lengths identical. Therefore, without losing any generality we stick to the case where $a_1=a_2=a_3=a_4=a$ and present the results in the following section.
\begin{figure}
\centering \figspace
\centerline{\includegraphics[width=0.95\columnwidth]{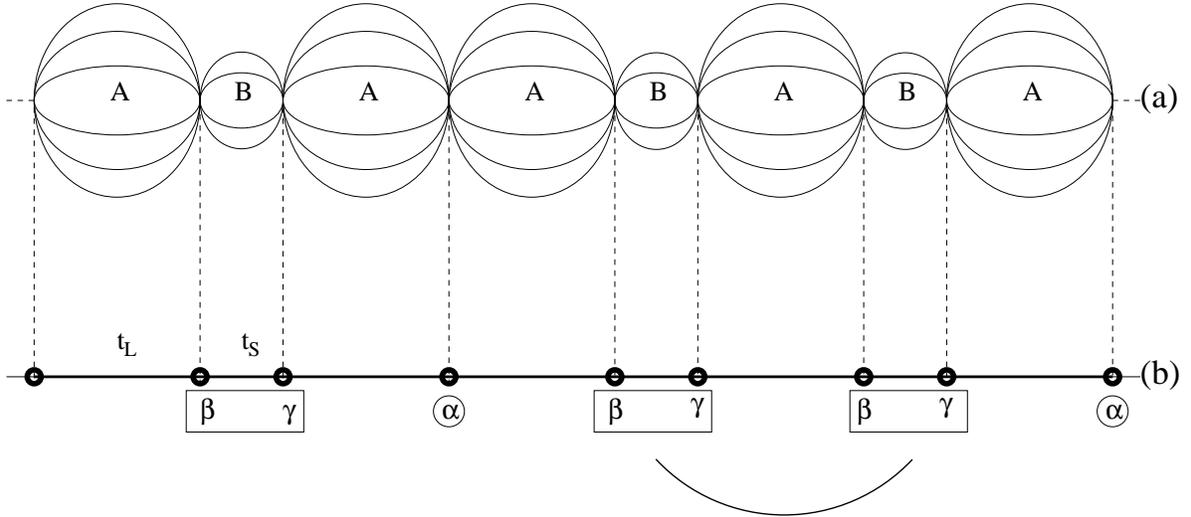}}
\caption{\label{fig1}
(a) A portion of an infinite Fibonacci array of waveguide loops.
(b) The equivalent 1d Fibonacci chain. The smallest resonating clusters
$\beta\gamma$ and $\alpha$ have been marked.} 
\end{figure}
\section{results and discussion}
\subsection{Band Structure}
\begin{figure}
\centering \figspace
\centerline{\includegraphics[width=0.95\columnwidth]{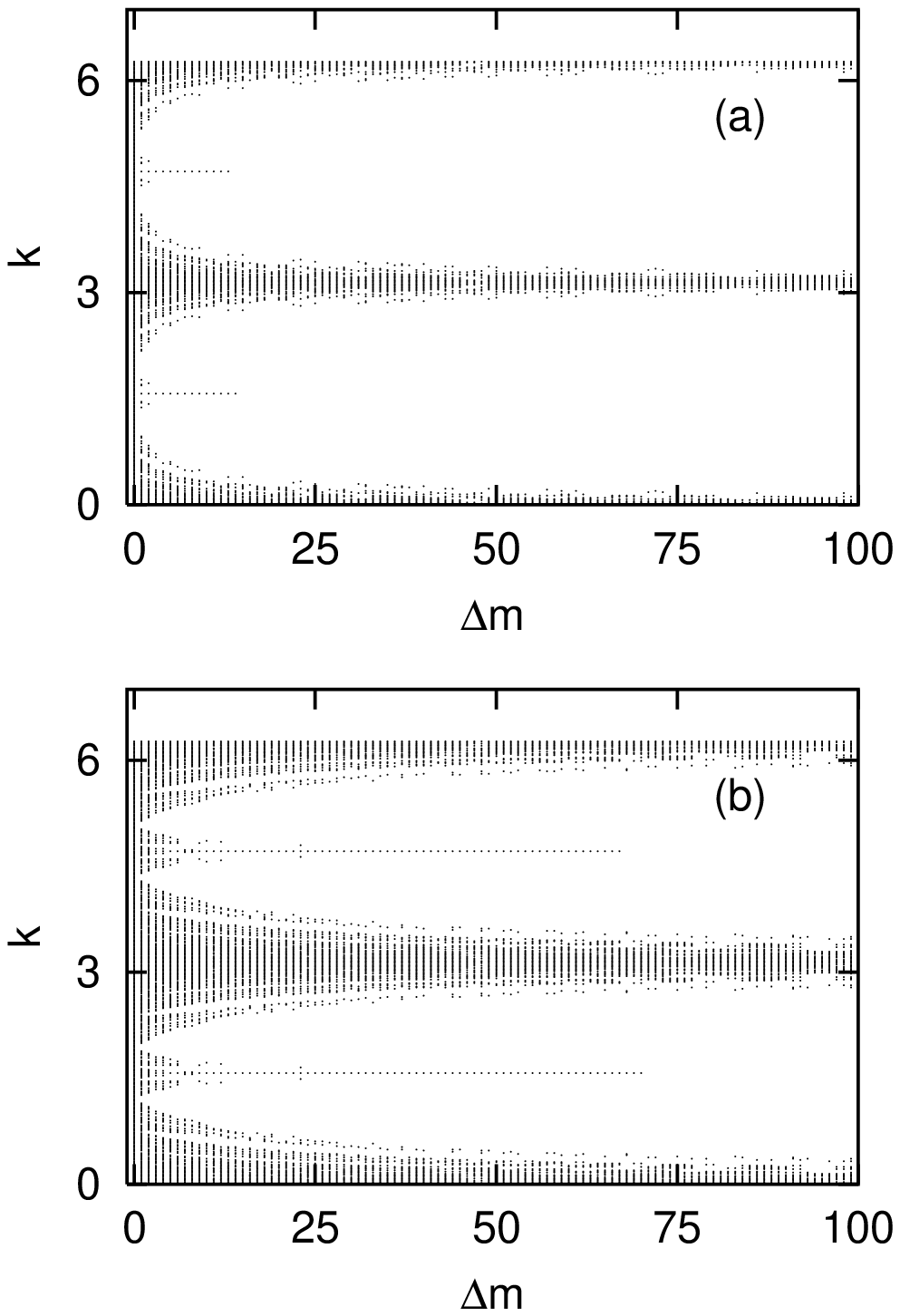}}
\caption{\label{fig2}
Allowed wave vectors for a $55$-loop system as a function of $m$ with,  
 (a) $n = 1$, and (b) $n = 5$.}
\end{figure}
The first few generation in a Fibonacci sequence of $A$ and $B$ are as follows: $S_1=A$; $S_2=AB$; $S_3=ABA$; $S_4=ABAAB$ and so on. An infinite Fibonacci sequence of loops $A$ and $B$ (or of $t_L$ and $t_S$) may be recursively generated by periodically repeating rational approximants of increasing size. By noting the individual sites (in terms of the equivalent one-dimensional chain), one can calculate the product transfer matrix across an $l^{th}$ generation Fibonacci rational approximant. The condition for a certain k-value to be an ``allowed'' one is, given by $\mid x_l\mid \le 1$, where, $x_l=\frac{1}{2}TrM_l$ \cite{mk83}, $M_l$ being the transfer matrix for the $l^{th}$ generation Fibonacci lattice, and satisfies the recursion relation $M_l=M_{l-2}M_{l-1}$ with $M_1=M_\alpha$ and $M_2=M_{\gamma \beta}=M_\gamma M_\beta$. The relation between $x_l$'s is,
\begin{displaymath}
x_l=x_{l-2}x_{l-1}-x_{l-3}, l\ge 4
\end{displaymath}
This is called the `trace map'. Using the trace-map method \cite{mk83} we have studied the distribution of the allowed values of the wave-vector $k$ as a function of $\Delta{m}=m-n$ keeping the lengths of all the segments in each of the loops $A$ and $B$ equal. For this we have selected an arbitrary rational approximant of a Fibonacci lattice, which consists of 55 loops. The case $\Delta m=0$ corresponds to the perfectly periodic arrangement of identical loops, for which we have a continuous distribution of ``allowed'' eigenvalues from $k=0$ to $2\pi$ (as we have shown) with $a=1$ in arbitrary units. With $n=1$ [Fig.2(a)] the spectrum starts splitting into three major sub-bands, with a couple of minor clusters of allowed k-values peeping in between the major sub-bands. These clusters do not grow when $\Delta m$ exceeds a certain value. Each major sub-band gets split up, exhibiting a fragmented structure as $m$ increases keeping $n$ fixed at 1. However, when $\Delta m$ becomes quite large, the further fragmentation of the major sub-bands becomes hard to detect numerically.
The features are sensitive to the starting value of $n$. Fig.2(b) shows the band structure, where the number of segments in the $B$-loop is $2n=10$. Here, the minor (weak) sub-clusters around $k=1.5$ and k close to 5 grow in width, as well as in length. With increasing values of $n$ the spectrum assumes a very prominent four-sub-band structure. The self-similar splitting of the spectrum is there as $\Delta m$ increases, which of course is limited by the resolution of the machine as $\Delta m$ increases to large enough values.  
\subsection{Transmission coefficient}
For computation of the transmission coefficient, we adopt the widely used technique \cite{ads81} of fixing semi-infinite leads to the extreme nodes of the equivalent one-dimensional Fibonacci lattice obtained by discretizing the wave equation. The lead is described by identical and equispaced sites with on-site potential $\epsilon_0$, set equal to zero and connected to each other via nearest neighbour hopping integral $t_0$, which we set equal to unity. An equivalent problem of an electron with energy $E=2\cos ka$ (in unit of $t_0$) travelling along such a lead and entering the finite sized sample will give us the transmittance of the actual problem of classical waves propagating through the Fibonacci SLS. Following Ref. \cite{ads81} the transmission coefficient of a finite equivalent one-dimensional Fibonacci segment consisting of sites $\alpha$, $\beta$ and $\gamma$ and hopping integrals $t_L$, $t_S$ is given by,
\begin{eqnarray}
T & = & \frac{4\sin^2\theta}{[(P_{12}-P_{21}+(P_{11}-P_{22})\cos \theta)^2+
(P_{11}+P_{22})^2\sin^2\theta]}
\end{eqnarray}
\noindent
where $P=\prod_i M_i$. $M_i$ are the transfer matrices corresponding to each site (including the boundary sites) in the equivalent chain that is clamped between the leads. Here $\theta=ka$.
 We have used the formulation first to get an idea of wave propagation through a periodic array  of $N$ identical loops, each consisting of the same `$m$' number of segments of equal length. The transmission coefficient for $N$ such loops in periodic arrangement turns out to be,
\begin{equation}
  T = \frac{16 m^2\sin^2 \theta}{[2m(U_{N-2}(x)-U_{N}(x))+\frac{(1-4m^2)}{2}U_{N-1}(x)\sin 2\theta]^2 +
    (1+4m^2)^2U_{N-1}^2(x) \sin^4\theta},
\label{eq-transmission}
\end{equation}
\noindent
where, $U_N(x)$ is the $N$th order Chebyshev polynomial of second kind, $a=a_1=a_2=a_3=a_4$ and $x=-\cos ka=-\cos \theta$.
\vskip .2in
In Fig.3 we plot the transmission coefficient for periodic array of 55 identical loops. It is interesting to see that in the present model the loops touch each other, and there is no real gap opening up in the spectrum, except for the isolated zeroes at $ka=0,\pi,2\pi,....$. The envelope of the values of $T$ between consecutive zeroes can be worked out from (9) to drop following a power law, viz, $T \sim \frac{1}{m^2}$ with increasing values of $m$, when $m$ is large.
\begin{figure}
 \centering \figspace
 \centerline{\includegraphics[width=0.95\columnwidth]{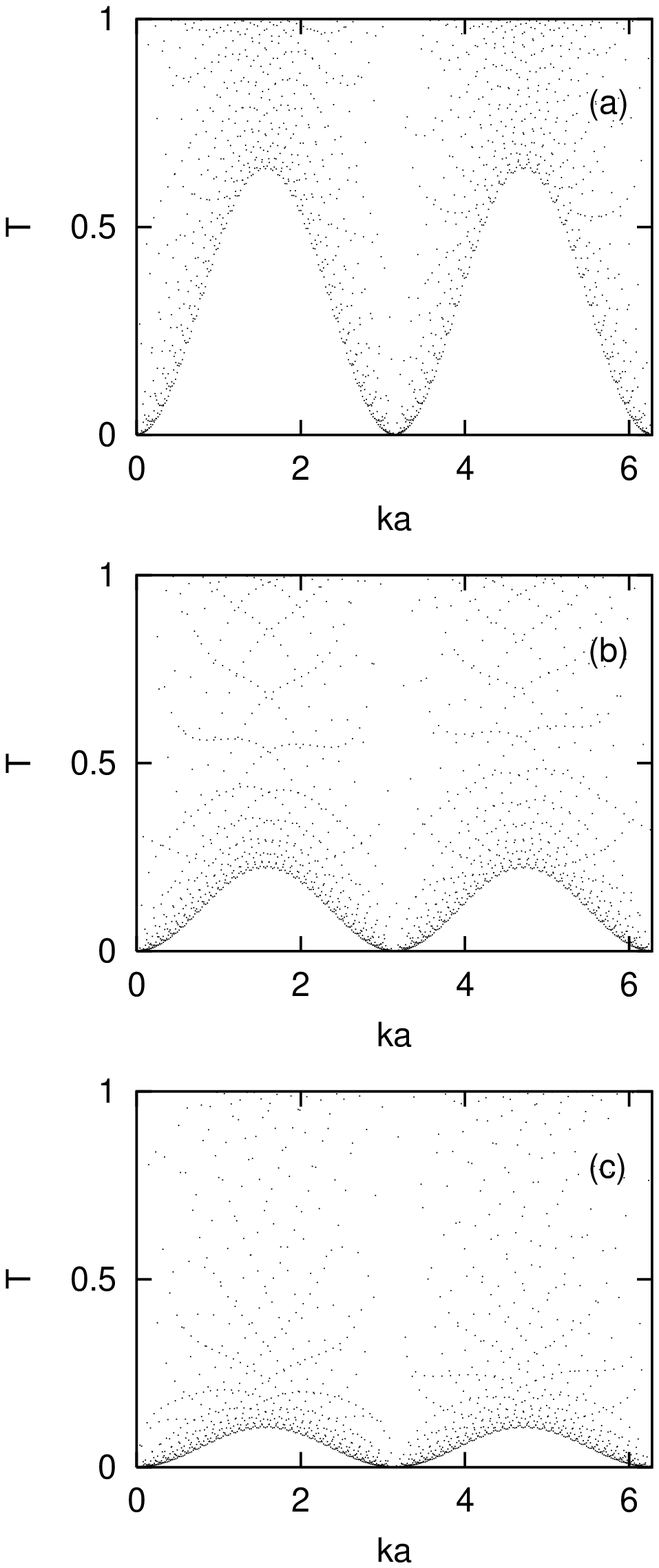}}
\caption{\label{fig3}
 Transmission coefficient $T$ versus $ka$ for a periodic array of
$55$ identical loops for (a) $m = n = 1$, (b) $m = n = 2$, and 
(c) $m = n = 3$. We have chosen $a_1 = a_2 = a_3 = a_4 = a = 1$ in 
arbitrary unit.}
\end{figure}
\vskip .2in
The calculation of transmission coefficient for a Fibonacci array of two different loops brings out an interesting difference. Fig.4 shows the transmission coefficient against the wave-vector for a 9th generation Fibonacci sequence of 55 loops. The loops differ only in the values of the number of waveguide segments $m$ and $n$, $n$ being fixed at 1. We have kept $a_1=a_2=a_3=a_4=a$ as before, and results are shown for $0\le ka \le 2\pi$ with $m=2$ [Fig.4(a)]. The corresponding lengths of the wave-guide segments with frequency range for an electromagnetic wave for example, can easily be extracted to design a possible network for the light localization in a quasi-periodic network. The gaps in the spectrum are quite clear. With $m$ taking up values 4 [Fig.4(b)] and 6 [Fig.4(c)], the gaps clearly widen and the transmission spectrum exhibits a global three sub-band structure. It should be appreciated that the transmission spectrum is sensitive to the absolute values of the parameters of the system. For example, in any given generation, by increasing the lengths of the segments for any loop $A$, or $B$, or by increasing the value of $n$ for a fixed $m$ ($>n$) one finds the growth of new sub-bands in the transmission characteristics. However, the additional sub-bands disappear when one increases the difference between $m$ and $n$ appreciably. 
\begin{figure}
 \centering \figspace
 \centerline{\includegraphics[width=0.95\columnwidth]{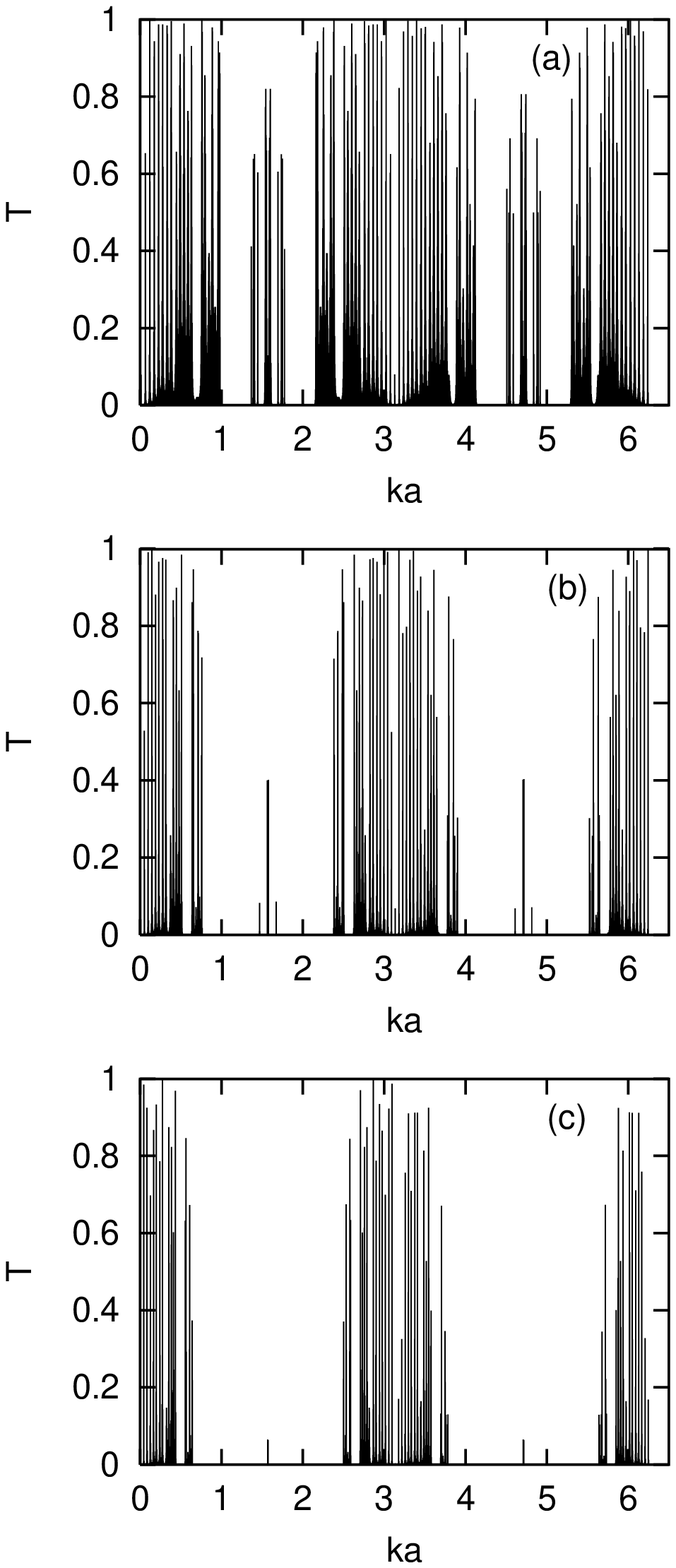}}
\caption{\label{fig4}
 Transmission coefficient $T$ versus $ka$ for a Fibonacci array of $55$ loops.
 The two different loops $A$ and $B$ are characterized by (a) $m = 2$, $n = 1$,
(b) $ m = 4$, $n = 1$, and (c) $m = 6$, $n = 1$. The lengths of the segments
are as in Fig.3.}
\end{figure}
\vskip .2in
For certain values of the wave-vector, appropriate choice of the lengths of the segments may lead to interesting fluctuations in the transmission spectrum. This behaviour is, however, sensitive to the number of $A$ and $B$ loops in the Fibonacci array. To clarify, let us define $\theta_j=ka_j$, $j=$1,2,3 and 4. It is simple to work out that, for $\theta_1=\frac{\pi}{2}$ and $\theta_3=\frac{\pi}{6}$, with $\theta_1+\theta_2=(2p+1)\pi$ and $\theta_3+\theta_4=(2q+1)\pi$, $p$ and $q$ being integers, each of the transfer matrices $M_\alpha$, $M_\beta$ and $M_\gamma$ becomes equal to $-i\sigma_y$, provided one adjusts $m$ and $n$ such that, 
\begin{equation}
m\sin\theta_3=n\sin\theta_1.
\end{equation}
In the equivalent electron problem we now have $E=2\cos\theta_1=0$ and $\epsilon_i=0$ with $t_L=t_S$. It implies that the equivalent one-dimensional chain becomes indistinguishable from a perfectly ordered chain of pseudo-atoms. In this situation, a straightforward calculation shows that, a Fibonacci array of loops at the $3l$ and $(3l+1)^{th}$ generation, with $l\ge 1$, will exhibit a $\frac{1}{m^2}$ decay in the transmission coefficient. For all other generations, with the above specification of parameters, we get $T=1$.
\begin{figure}
 \centering \figspace
 \centerline{\includegraphics[width=0.95\columnwidth]{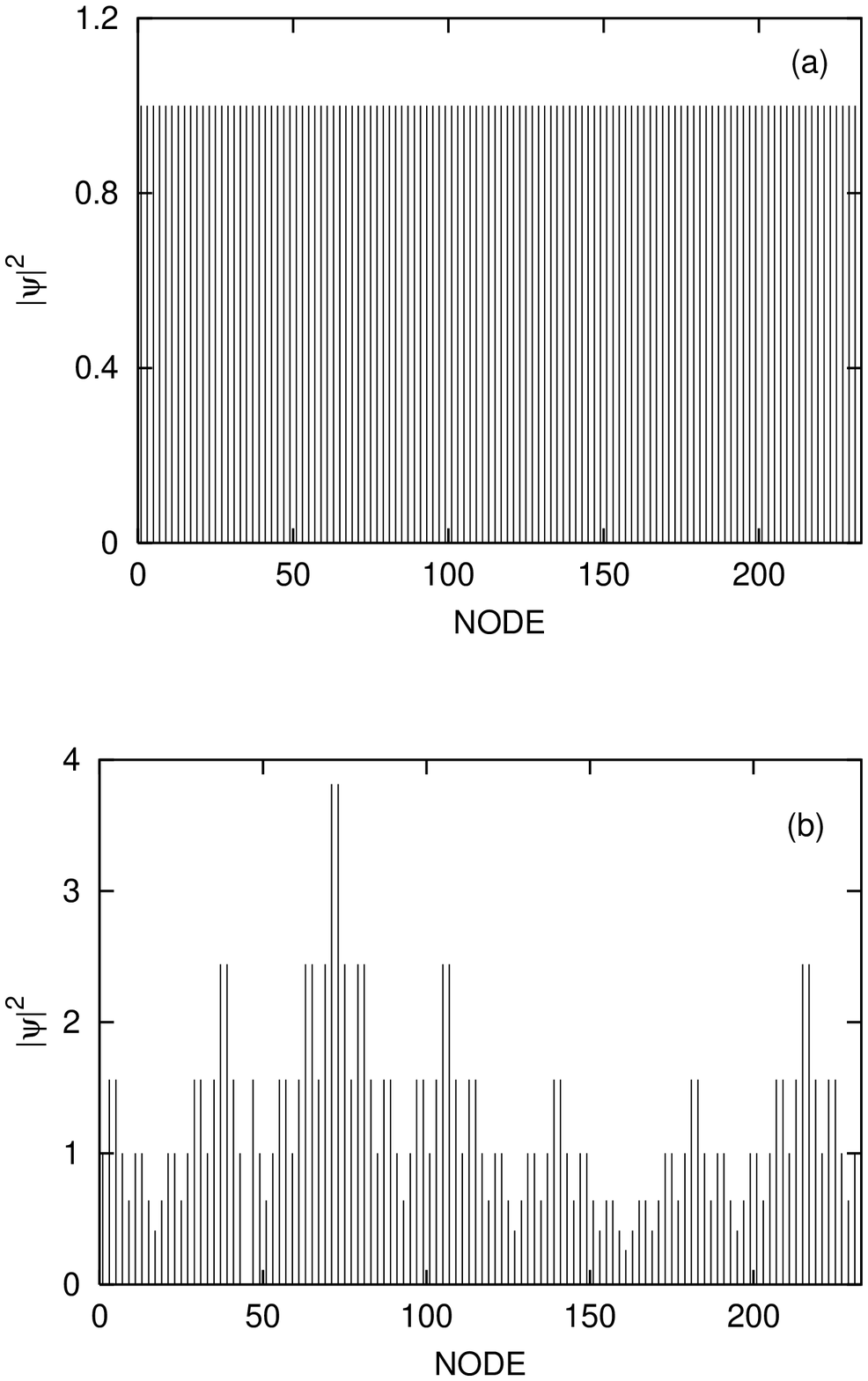}}
\caption{\label{fig5}
Intensity distribution at the nodes of a Fibonacci waveguide array consisting of $233$ loops with (a) $m=2n$, $n=1,2,.$ for a periodic distribution, and (b) $m \ne 2n$ for a self-similar (fractal) distribution. In either case, we have chosen $\theta_1=\pi/2$, $\theta_2=5\pi/2$, $\theta_3=\pi/6$, and $\theta_4=29\pi/6$.}
\end{figure}
\subsection{Dimer like correlations and resonance}
Looking at the equivalent one dimensional Fibonacci chain (Fig.1b), we find that the cluster $\beta\gamma$ (put in a box) appears in pairs, as well as in isolation. The site $\alpha$ (put in a circle) however, is always single. The $\beta\gamma-\beta\gamma$ pairs, whenever they appear, are always flanked by two $\alpha$ sites. Now, if the traces of the $2 \times 2$ transfer matrices $M_{\gamma\beta}$ and $M_\alpha$ can be made to vanish simultaneously for a certain combination of the system-parameters and the wave vector, then the cluster $\alpha-\beta-\gamma-\beta-\gamma-\alpha$ will contribute an identity matrix to the entire product of the transfer matrices at any given generation of the Fibonacci chain \cite{sc02}. This leads to a {\it resonance} in the above cluster. As has been explained elsewhere \cite{sc02}, such resonance then takes place locally throughout the chain, reducing the long product of transfer matrices, viz, $M_{l+6}$ to $M_l$ for $l \ge 1$. This also explains the six cyclic invariance of the transfer matrices in a Fibonacci chain \cite{sc02}. However, the sizes of the minimal clusters responsible for resonance depend on the choice of the wave vector, and definitely, on the combination of the other parameters of the system. To clarify, we work out a specific example. Let us select, $\theta_1+\theta_2=(2p+1)\pi$ and $\theta_3+\theta_4=(2l+1)\pi$, $p$ and $q$ being integers, equal or unequal. This choice automatically ensures $\epsilon_\alpha=\epsilon_\beta=\epsilon_\gamma=E$ and a six-cycle of the transfer matrices for all $k$-values consistent with the above conditions for any suitable combination of the $a_i$'s \cite{sc02}. In particular, when the special condition (10) is satisfied, it is simple to check that $t_L=t_S$ and $M_\alpha=M_\beta=M_\gamma=-i\sigma_y$. The products $M_{\gamma\beta}^2=M_{\alpha}^2=-I$, and the resonance condition, as discussed above, is trivially satisfied. With $\beta\gamma-\beta\gamma$ and $\alpha-\alpha$ playing the roles of the minimal clusters responsible for resonance. Otherwise, if $m\sin\theta_3 \ne n\sin\theta_1$ i.e. $t_L \ne t_S$, we get back the well known pure transfer model of a Fibonacci chain \cite{mk83} with $E=2\cos\theta_1$ at its band-centre. Again, a six-cycle wave-function results, and the intensity $\mid\psi_i\mid^2$ has a purely multi-fractal character \cite{mk83}. The multifractality can be removed by controlling the values of $m$ and $n$. For example, with $ka_1=\frac{\pi}{2}$ and $ka_3=\frac{\pi}{6}$, we get $t_L=t_S\frac{m}{2n}$. So, $m=2n$ makes the equivalent one-dimensional lattice completely periodic. $\mid\psi_i\mid^2$ is also periodic in this case, though the loops are arranged in a Fibonacci sequence. We get an extended eigenmode. For all other cases with $m=2n$, the transfer model, and hence the fractal nature of $\mid \psi_i \mid^2$ at the nodes is restored. This feature is illustrated in Fig.5.       
\section{conclusion}
We investigate the propagation of classical waves in a Fibonacci array of waveguide network. For specific considerations of the system parameters one can have extended, non-fractal wave function is the system, contrary to the quantum counterpart of the problem. Resonant transmission in such network can be attributed dimer-like correlation in the placement of the loops. The study shows that one can design such a network as an example of quasi-periodic photonic band gap system.
\section{acknowledgement}
AC thanks CSIR, India for financial assistance through grant no. 03(0944)/02/EMR-II. 


\begin{thebibliography}{10}
\bibitem{cms95}see for example, {\it Photonic band gap materials}, edited by C. M. Soukoulis (Kluwer, Dordrecht, 1995).
\bibitem{sj87}S. John, Phys. Rev. Lett. 58 (1987) 2486.
\bibitem{ey87}E. Yablonoviteh, Phys. Rev. Lett. 58 (1987) 2059. 
\bibitem{msk96}M. S. Schwa, Int. J. Mod. Phys. B 10 (1996) 977.
\bibitem{dzz94}D. Z. Zhang, W. Wu, Y. L. Zhang, L. Li, B. Y. Cheng and G. Z. Yang, Phys. Rev. B 50 (1994) 9810.
\bibitem{slm91}S. L. Macall, P. M. Platzman, R. Dalichaouch, D. Smith and S. Schultz, Phys. Rev. Lett. 67 (1991) 2017; Nature (London) 354 (1991) 53. 
\bibitem{dsw97}D. S. Wiersma, P. Bartolini, Ad. Lagendijk and R. Righini, Nature (London) 390 (1997) 671.
\bibitem{zqz98}Z. Q. Zhang, C. C. Wong, K. K. Fung, Y. L. Ho, W. L. Chan, S. C. Kar, T. L. Chan and N. Cheung, Phys. Rev. Lett. 81 (1998) 5540.
\bibitem{jov99}J. O. Vesseur, B. Djafari-Rouhani, L. Dobrzynski, A. Akjouj and J. Zemmouri, Phys. Rev. B 59 (1999) 13446.
\bibitem{aa03}A. Mir, A. Akjouj, J. O. Vesseur, B. Djafari-Rouhani, N. Fethouhi, E. H. El Boudouti, L. Dobrzynski and J. Zemmouri, J. Phys.: Condens. Matt. 15 (2003) 1593.  
\bibitem{aa04}A. Akjouj, H. Al-Wahsh, B. Sylla and B. Djafari-Rouhani, J. Phys.:Condens. Matt. 16 (2004) 37.  
\bibitem{zqz94}Z. Q. Zhang and P. Sheng, Phys. Rev. B 49 (1994) 83.
\bibitem{mk83}M. Kohmoto, L. P. Kadanoff and C. Tang, Phys. Rev. Lett. 50 (1983) 1879; M. Kohmoto, B. Sutherland and C. Tang, Phys. Rev. B 35 (1987) 1020.
\bibitem{ss93}S. Sil, S. N. Karmakar, R. K. Moitra and A. Chakrabarti, Phys. Rev. B 48 (1993) 4192.
\bibitem{dhd90}D. H. Dunlop, H. L. Wu and P. W. Phillips, Phys. Rev. Lett. 65 (1990) 88; D. H. Dunlop and P. W. Phillips, J. Chem. Phys. 92 (1990) 6093.
\bibitem{kk93}P. K. Dutta, D. Giri and K. Kundu, Phys. Rev. B 47 (1993) 10727; J. C. Flores and M. Hilke, J. Phys. A: Math. Gen. 26 (1993) L1255.
\bibitem{ads81}A. Douglas Stone, J. D. Joannopoulos and D. J. Chadi, Phys. Rev. B 24 (1981) 5583.
\bibitem{sc02}S. Chattopadhyay and A. Chakrabarti, Phys. Rev. B 65 (2002) 184204.
\end{thebibliography}
\end{document}